\documentstyle[12pt,procsla]{article}
  \catcode`\@=11
  \long\def\@makefntext#1{
  \protect\noindent \hbox to 3.2pt {\hskip-.9pt  
  $^{{\ninerm\@thefnmark}}$\hfil}#1\hfill}		
  
  \def\@makefnmark{\hbox to 0pt{$^{\@thefnmark}$\hss}}  
  \def\applt{\mathrel{\hbox{\rlap{\hbox{\lower4pt\hbox{$\sim$}}}\hbox{$<$}}}}
  \def\appgt{\mathrel{\hbox{\rlap{\hbox{\lower4pt\hbox{$\sim$}}}\hbox{$>$}}}}
  	
  \def\ps@myheadings{\let\@mkboth\@gobbletwo
  \def\@oddhead{\hbox{}
  \rightmark\hfil\ninerm\thepage}   
  \def\@oddfoot{}\def\@evenhead{\ninerm\thepage\hfil
  \leftmark\hbox{}}\def\@evenfoot{}
  \def\sectionmark##1{}\def\subsectionmark##1{}}
  
\begin{document}
  
  \centerline{\normalsize\bf THE ULTRA HIGH ENERGY NEUTRINO--NUCLEON CROSS
  SECTION
  \footnote{Conf. Proc. 7$^{\rm th}$ International Symposium on
  Neutrino Telescopes, Feb. 27 - Mar. 1, 1996.}} 

  \vspace*{0.6cm}
  \centerline{\footnotesize  JOHN P. RALSTON and DOUGLAS W. MCKAY}
  \baselineskip=13pt
  \centerline{\footnotesize\it  Department of Physics \& Astronomy, University
  of Kansas}
  \baselineskip=12pt
  \centerline{\footnotesize\it Lawrence, Kansas 66045, USA}
  \centerline{\footnotesize E-mail: ralston@kuphsx.phsx.ukans.edu}
  \centerline{\footnotesize E-mail: mckay@kuphsx.phsx.ukans.edu}
  \vspace*{0.3cm}
  \centerline{\footnotesize and}
  \vspace*{0.3cm}
  \centerline{\footnotesize GEORGE M. FRICHTER}
  \baselineskip=13pt
  \centerline{\footnotesize\it Bartol Institute, University of Delaware,
  Newark, DE 19716 USA}
  \centerline{\footnotesize E-mail: frichter@lepton.bartol.udel.edu}  
  \vspace*{0.9cm}
  \abstracts{The ultra-high energy neutrino nucleon cross section grows at a
surprising rate with energy due to QCD effects in the target.  Recent
electroproduction data allows an update of earlier predictions.
We compare the results of our own calculations with those of other groups, and
critically review the foundation and reliability of the calculations.  The question of the ``new
physics" potential of neutrino telescopes sensitive to the total cross section
in multi-PeV energy domain is considered.  We point out a loophole in the 
arguments which might be an important consideration in
extrapolating the cross section to extremely high energies. 
}

  \normalsize\baselineskip=15pt
  \setcounter{footnote}{0}
  \renewcommand{\thefootnote}{\alph{footnote}}
  \section{Overview}

	The neutrino-nucleon interaction cross section illustrates 
different physics in different energy regimes.  In the ultra-low 
energy region $\sigma_{tot}$ goes like $E_\nu^2$, an effect of the
non-relativistic phase space of the final state electron.  When the  
reaction products are relativistic, $\sigma_{tot}$ goes like $E_\nu$, which 
can be guessed from dimensional analysis of the 4-Fermi theory:  
$\sigma_{tot}( s = 2 m_N E_\nu)\cong G_F^2 s$, where $M_N$ is the target
nucleon mass.  In both of these regimes the cross section is exceedingly small:
a  
parameterization\cite{PDG} of the 4-Fermi region which agrees with data is 
\begin{equation}
\sigma_{tot,4-Fermi}^{\nu N}  = 6.7 \times 10^{-39} cm^2 (E_\nu/GeV). 
\end{equation}

	Despite the smallness of the numbers, the linear increase of 
$\sigma_{tot}$ with energy indicates a problem with probability 
conservation, the well-known ``unitarity violation" of the 4-Fermi 
model.  The problem is fixed by including $W$-propagator effects 
present in the Standard Model,  which kills the linear rise
with  
energy at about $E_\nu = m_N/m^2_W$.  If only the effects of the $W$-propagator
on 
an elementary target were included, then the total  
cross section would only vary logarithmically with energy above 
this region. 

	The story becomes more interesting, especially for neutrino 
telescope purposes, in the ultra-high energy region.  This is because 
the nucleon target is a complicated composite object, full of quarks,
antiquarks and gluons.  To the fast moving $W$ boson the proton looks 
sort of like a high-octane {\it grappa}~\cite{grappa}  of
quark-antiquark pairs. More formally, the reaction is best described 
using the Bjorken variable $x_{Bj} = Q^2/ (2 m_N E_W)$ and the momentum
transfer $Q^2$  of 
deeply inelastic scattering. With a spatial resolution set by $Q^2$, the 
$W$ sees quarks (and antiquarks) which are carrying momentum 
fraction $x =x_{Bj}$.  The total cross section goes like each elementary 
cross section, times the proper number of quarks and antiquarks and 
their known weak charges-squared.

	Recent experimental studies at HERA have confirmed  
that there are surprisingly many 
quark-antiquark pairs in the nucleon.  This QCD phenomenon builds a
significant enhancement of an electroweak cross section; it is a 
big effect that, if neglected, would lead to sizable errors in the 
energy region above $E_\nu = 100$ TeV.  The effects have been 
incorporated~\cite{fmr} with calculations in the Standard Model, resulting in 
handy, reliable formulas to be displayed below.  From the neutrino 
telescope point of view, the QCD enhancement is very welcome, 
boosting the detection rates for interesting cosmic ray neutrinos 
from ``out there" as we study them ``down here" on planet Earth.

\section{The Calculation}

	The effects of $W-$ and $Z-$ exchange in the Standard Model are 
described by well understood prefactors multiplying certain 
structure functions ${\cal F}^{\nu N}$ that parameterize the target.  The
behavior of  
neutrino versus anti-neutrino beams and  the $y$-dependence of
differential cross sections are fixed by boson exchange 
kinematics and the known mixtures of vector and axial currents. Due to the
special kinematic conditions,  
the differential cross sections at UHE can be written directly as 
\begin{equation}
{d\sigma\over dxdy}\bigg|_{UHE}={G_F^2M_W^2\over 2\pi}{M_W^2\over
s}(1-y+y^2/2){\cal F}^{\nu
N}_2(x,Q^2=sxy)/(M^2_W/s+xy)^2
\end{equation}

A useful rule of thumb tells what 
the dominant $Q^2$ and $x$ regions in the integrations will be. One needs 
 $Q^2 \applt M_W^2\applt 6.6 \times 10^3$ GeV$^2$.  In 
addition, the important part of the reactions come when the 
squared center of mass energy of the struck--quark and neutrino 
subsystem is big: 
\begin{equation}
2 M_N E_\nu  x\appgt M^2_W.
\end{equation}              
One therefore needs the 
quark and anti-quark number densities $(q_i(x)$, $\bar q_i(x) )$  in the 
region $x \appgt M^2_W/2  E_\nu$, or in simpler form, $x = 3.7 \times 10^{-3} 
(E_\nu/$ $100 TeV)^{(-1)}$.  The first group to have noticed  that small-$x$
is 
important seems to be Andreev, Berezinsky and Smirnov~\cite{ABS}.  Those
early  
calculations used  parameterizations of the quark 
structure functions which were no good at small-$x$.  Our contributions 
began with Ref [4], where we put in more relevant QCD-inspired 
small--$x$ structure, also studied again by Ref. [6].

	The H1 and ZEUS experimental groups~\cite{data} have recently found that the 
relevant number of partons involved in electroproduction, $e^2_i x ( 
\bar q_i(x) +q_i(x) )$, increases like a fractional power of $1/x$.  
Previously data did not exist for the region $x\applt 10^{-2}$; HERA extends
measurements down to $x =10^{-4}  - 10^{-5}$.  However, the $Q^2$ values 
are mismatched for the cross section we need here:  HERA has $Q^2$ of 
order 1-10 GeV$^2$ at smallest $x$. Moreover, 
the particular mix of flavors and charges measured in the 
experiment is not the same as the one involved in neutrino (or anti-neutrino)
scattering.   

	We can overcome these problems.  In the small-$x$ region of $x\applt
10^{-2}$, the proton loses almost all identity, becoming almost pure 
flavor singlet, with equal numbers of quarks and antiquarks.  This is 
called the ``sea dominance". We know this is true by studying the 
evolution in $x$ and $Q^2$, which is dominated by gluons making gluons, 
and then coupling to quark and anti-quark pairs in the sea.  Because 
of this, one can safely transcribe data measured by the photon into 
that measured by $W$'s and $Z$'s by suitable changes of coupling 
constants, as described in Refs.[3,4].  This approximation, also called
``neglecting the valence", is the right thing to do to understand the 
physics.  It has been confirmed to be a good approximation at UHE~\cite{sea}.

 	  The $Q^2$ evolution based on these ideas describes the data 
beautifully: our updated results for the QCD evolution\cite{fmr} subsumes several 
dozen data points with $\chi^2/$degree of freedom =0.8.  The theoretical
uncertainty from this procedure is typical of next-to-leading order 
QCD: it is about 10\% relative error.  This can be confirmed by 
examining the results of different fits which select different sets 
of data in global analyses of structure functions.~\cite{MRS} Our own
calculations~\cite{fmr} are unique in using theoretical 
approximations which fit the small-$x$ region locally.  They have 
recently been updated by new data at even smaller $x$, slightly 
changing the fits to give new results reported below. The fact that
perturbation QCD reproduces the $Q^2$ dependence so well strongly indicates that
the theory is reliable. As a point of 
caution, present theory remains limited to about 10\% accuracy by a host of
uncertainties, so that at this point citing 3--digit accuracy in numerical work
seems unjustified. 

	After fitting the data at low $Q^2$, we use QCD evolution to 
predict the parton distributions measured by an UHE neutrino. The 
singular nature of the newly measured parton distributions makes 
our conservative estimates of some years ago\cite{{RM86},{QRW}} obsolete.  In addition, the
objects  
known as dGLAP splitting functions\cite{dGLAP} must be employed at both leading 
and next-to-leading order.  Since we did this, and all groups which 
have parton distributions on the market also do this, the procedure 
is not controversial, and this part of the physics can be considered 
reliable.  

	Let us note for later reference that the smallest $x$ values 
actually measured puts an upper limit on the region where the cross 
section can be predicted with high certainty:  $x\geq 3 \times 10^{-5}$ 
translates into a region of $E_\nu \leq 10^4 TeV = 10 PeV$. This happens to
include the region where radio detection is the best prospect, as discussed
elsewhere in this volume~\cite{frichter} and in the literature.~\cite{radio}

\section{Results and Reliability}

	In the region of energy from 1 TeV to 10 TeV the 
approximation of neglecting the valence is not wonderful, but tolerable with
errors of 25--50\%.  The method begins to  
work well in the next decade of energy, and a single published data 
point\cite{point} at 70 TeV is nicely predicted.\cite{fmr}  Note that a
different number is used in Ref. [8], which cites unpublished values for
the average used in its comparison. In the region of
$E_\nu > 100$ TeV and on up, it is a  
useful approximation of our recent work to write simply
\begin{equation}
\sigma_{tot} = 1.2 \times 10^{-32} cm^2 ( E_\nu/ 10^{18} eV)^{0.40}
\end{equation}

	Resonant $W$ production from electron antineutrinos on electron 
targets (the ``Glashow resonance") is not included in the formula 
above.  The resonance occurs at $E_\nu = 6.7$ PeV, with an energy width 
of about $\pm 130$ TeV.  The resonance is tall but narrow on the 
cosmic ray scale, and it is not a major effect in the absorption of a 
broad neutrino spectrum. However, under certain circumstances the 
resonance might be useful to calibrate detection methods (including 
radio\cite{{radio},{thanks}}) in this region. 

	It may be surprising that particle physics can make reliable 
predictions for energies that are orders or magnitude higher than 
those measured in the laboratory.  It can done because ``new physics" 
is severely restricted by precision electroweak measurements.  
There is next to no room for bosons lighter than the known $W$ and $Z$, 
that couple leptons to quarks in the t-channel.  If one made the 
hypothesis anyway, the couplings would have to be so weak that the 
processes could not compete in the total cross section.  Of course 
there might be weak bosons that are much, much heavier than the $W$.  
This has caused colleagues to ask us whether the puzzles recently 
seen in Fermilab jet cross sections\cite{jets}, which may indicate new glue-type physics,
might not be replayed in some form for neutrinos. The  
practical  answer is no: the total cross section is quite insensitive 
to such effects which occur at kinematic boundaries.  For example, 
the proton total cross section is practically unaffected by  those anomalies: $\sigma_{tot}$ is still determined by soft 
physics and pion exchanges even at the gigantic energies of colliding Tevatron beams.  Similarly, new physics for neutrinos 
might show up in a rare process with very large momentum transfer, 
but such events are very unusual and require instruments specialized 
to see them.  Again, the Fermilab events may be explainable by small
adjustments in the parton distributions near $x\approx 1$. The effects of
these on $\sigma_{tot}^{\nu N}$ at UHE are very small. For the total cross section we cannot beat the vector 
bosons we already know; it is a general rule of scattering theory 
that total cross sections are dominated by the lightest t-channel 
exchange.  As for something new happening in the s-channel, leptons 
do not annihilate with quarks, up to exceptionally good limits from 
proton (non)- decay, etc. Speculative $s$--channel production of heavy bosons
on atomic electrons can be contemplated. This might make a big effect, but only
in a limited resonance  region. Could a neutrino telescope see this? Perhaps
the question merits investigation. 

Another possibility for finding new physics has been suggested by Learned and
Pakvasa\cite{LP}. They point out that tau neutrinos, if any, would be an
indication of neutrino oscillations, with a signal of a ``double--bang" at PeV
energies from the charged current event and later decay of the tau lepton.

	Our group originally got interested in the total cross section 
problem via a long series of efforts to evade our own  logic. The motivation
was the Cygnus $X3$ problem, which we hoped new neutrino physics might
explain.\cite{{RM86},{Cyg}}
 We only
discovered the small-$x$ QCD effect by going through  
all possibilities we could imagine, convincing ourselves that it is 
the biggest non-resonant thing that actually can happen, and finding Ref. [5] after 
working things out our own way.

\section{Third Order Second Thoughts}

	Having often reconsidered  our second thoughts, we were 
asked\cite{xthanx}  to think more deeply on these questions: is it 
really rock-hard-reliable that QCD and the Standard Model can 
predict the cross section?  There is a loophole.  It is in the extrapolations
presently used to go to smaller-$x$  
than has been measured.  As mentioned earlier, we know the $x$-distributions down
to about $10^{-5}$, corresponding to laboratory-based predictions up to
neutrino energies of about 10 PeV.   
Extrapolations above this energy are really based on theoretical 
expectations, which might be right or wrong. 

	Two approaches exist.  One is to apply the criteria of Gribov, 
Levin and Ryskin\cite{GLR}, which allows parton densities at small-$x$ to 
continue to grow until unitarity is confronted at the $Q^2$ of the 
process.  Theorists who try to predict the $x$-dependence\cite{smallx} tend to 
predict a power-law rise (which is an eigenvector 
of dGLAP evolution).  In this sense the theory is under control. For $Q^2 = 
M_W^2$, the results of this approach is continued growth of the cross 
section up to an extraordinarily high energy, of order $E_\nu\approx 10^{20}$ eV.  The 
point of leveling--off occurs where the $\nu p$ cross section 
approaches the $pp$ total cross section!  Since we first 
applied this criterion to the neutrino problem, we are fond of it and 
have promoted it accordingly.

	On the other hand, hadrons are complicated beasts that nobody 
understands well.  It would not violate any known laws of Nature for 
the observed small-$x$ growth simply to stop.  This might happen at a 
point earlier than current estimates indicate: experimental 
measurements are needed to know for sure.  A guesstimate might be  
$(\alpha_s/2 \pi) \log(1/x)\cong 1$  indicating saturation at $x$ below the 
$10^{-5}$ region.  If this is correct, then one cannot absolutely rely on 
having a gigantically rising cross section forever, and $\sigma^{\nu
N}_{tot}$ might round 
off  above 10 PeV.  As responsible theorists, we are 
obliged to note this disconcerting possibility. We believe that the value at 10
PeV serves as a lower bound on $\sigma^{\nu N}_{tot}$ if this phenomenon should
occur.

\section{Acknowledgments}

This work was supported by Department of Energy Grant 
Numbers DE FG02 85 ER 40214 and DE FG02 91 ER 40626.A007, 
and by the {\it Kansas Institute for 
Theoretical and Computational Science}.  
We thank several 
colleagues, including John Learned, Francis Halzen, Rocky Kolb, 
and Al Mann for helpful remarks.

  \end{document}